# Adaptable Plug and Play Security Operations Center Leveraging a Novel Programmable Plugin-based Intrusion Detection and Prevention System

Ahmed S. Shatnawi[a,*,1], Basheer Al-Duwairi[b,2], Mahmoud M. Almazari[b,3], Mohammad S. Alshakhatreh[b,4], Ahmad N. Khader[b,5] and Abdullah A. Abdullah[b,6]

[a]*Department of Software Engineering & Security, Jordan University of Science & Technology, Irbid, Jordan*
[b]*Department of Network Engineering, Jordan University of Science & Technology, Irbid, Jordan*

**ABSTRACT**

The number of cyber-attacks have substantially increased over the past decade resulting in huge organizational financial losses. Indeed, it is no longer a matter of "if" but "when" a security incident will take place. A Security Operations Center(SOC) adoption will help in the detection, identification, prevention, and resolution of issues before they end up causing extensive cyber-related damage. In this paper, our proposed framework is brought about to address the problem that current open-source SOC implementations are plagued with. These include lack of ability to be strengthened on the fly, slow development processes, and their ineptness for continuous timely updates. We, herein, propose a framework that would offer a fully automated open-source SOC deployment; otherwise dubbed, a "plug-and-play framework"; full horizontal scalability incorporating a modular architecture. These underpinning features are meant to mitigate underlying SOC challenges, which often emerge as a result of many pre-determined and repeated processes, bolstering their ability for expansion with new tools. This is on top of enhancing their ability to handle more servers in the clusters as a single logical unit. We also introduce a new system of its kind called a Programmable Plugin-based Intrusion Detection and Prevention System (PPIDPS). This system will extend a SOC's ability to add any tool to the monitored devices while collecting logs that can trigger alerts whenever a suspicious behavior is detected.

## 1. Introduction

The number of security-threatening attacks has increased substantially in recent years; the expense of these attacks on companies has increased drastically [12]. Every organization needs improvements in security incident detection and prevention through continuous monitoring, data analysis, and oversight of network activities. Such security improvements can be achieved leveraging a very well structured methodology renowned as a SOC center; a robust defensive strategy recognized throughout the field of Cybersecurity. Recently, attacks including Advanced Persistent Threats (APT) that simultaneously utilize a varied range of attack vectors and patterns have surfaced [14]. Thus, their detection by point solutions is almost an unachievable task; these kinds of attacks are well organized, directed, financed, silent, unexpected, and very patient. Indeed, it is not anymore a matter of "if" but "when" a security incident will take place. Recent events are manifested in the massive breach that took place in the U.S during late 2020, where many firms including the US Departments of Treasury and of homeland security, state and defense are known to have been targeted. Around 58 percent of the healthcare sector in many countries experienced one form or another of a cyber-attack last year [39].

To deal with such situations, government agencies and industries can get empowered by readily installable and easily manageable SOCs adaptable to various organizational budgets. What currently exists on the market to help mitigate such cyber-crimes and help companies and organizations monitor and manage all aspects of their security concerns in real-time, either in the form of commercial products, i.e., SOC (Security Operations Center) usually costing them a fortune, or open source products that would burden the IT people handling these open source products. It is, therefore, fairly easy for SOC managers to get caught up by the false positive alerts, and be overwhelmed with the bombardment of buzzing configuration errors. In all, these considerations will lead to scenarios in which SOC administrators and other security concerned leaders would lose perspective on what actually matters, including knowledge of their risk profiles, taking the proper procedures in the face of threats, and building up systems and mechanisms to resolve severe security problems.

Machines are excellent at quickly performing redundant and laborious tasks, as manual human interactions commonly incur devastating consequences commensurate with the errors that can come about in the process. Consequently, such tasks should not be performed by the security analysts themselves. Automating the inner workings of low hanging fruit spontaneously allows security experts to focus more on context-essential tasks. Automation of a SOC ensures that both the acting staff and computing needs can be appropriately stream-

*Ahmed S. Shatnawi

✉ ahmedshatnawi@just.edu.jo (A.S. Shatnawi); basheer@just.edu.jo (B. Al-Duwairi); mahmoudalmazari@gmail.com (M.M. Almazari); hmodaa.sha@hotmail.com (M.S. Alshakhatreh); mahmoudalmazari@gmail.com (A.N. Khader); abdullahanashaj@gmail.com (A.A. Abdullah)
ORCID(s):

lined. Firmly aligned, the personnel and tools will do better and keep frequently encountered critical mistakes to a minimum. This may just be the absent piece in the whole puzzle that enables a SOC to be more proactive.

Furthermore, the proposed SOC architecture is one that would offer adaptability. Towards that end, different institutions have different requirements, risk profiles, departments, technologies, and budgets, all of which are essential considerations when designing a SOC architecture. Hence, a full-fledged assessment with reports from the organizational security experts regarding the risk profiles must be made. This will eventually allow the SOC to handle any risk and implement any countermeasures in the face of any threat to the SOC system. A method has yet to be found that makes network engineers and system administrators more capable of updating their intrusion detection techniques to mitigate the more progressively persistent attacks that may rely on zero-day vulnerabilities. Therefore, we propose a technique to enable any kind of code to be executed and gather any type of information to log them into the SOC center to enhance the intrusion detection or attack prevention methods. The technique can more incessantly monitor the end devices to detect anomalous behaviors and offer an intrusion prevention mechanism in due time. In this manner, our objective is to add some level of automation to the SOC, releasing some essential time to be invested more efficiently in securing the system.

This paper addresses a robust implementation of a fully automated SOC deployment, offering adaptability in handling the diversity of risk profiles of different organizations. Moreover, the proposed framework adds new security functionality to existing Open-Source Security Information and Event Management (SIEM) implementations. This, inherently will allow SOC managers to feel more confident about having a better leverage in securing the devices their teams are constantly monitoring.

The main contributions of this paper are as follows:

1. A new dynamic architecture framework that allows a security operations center's deployment on the fly, enabling horizontal scalability and expanding the security operations center toolchain.
2. A novel security operations center subsystem to adapt to different risk profiles and threats.
3. A new, a one-of-a-kind, framework that enables threat mitigation and collaboration between different security researchers and SOC teams around the globe by seamless SOC integration through a business communication platform.

This paper is organized as follows. In Section 2, we present background information and survey earlier research around the topic. Section 3 elaborates the analysis and design of the proposed solution. Section 4 reveals details of the proposed framework implementation; In Section 5, we discuss our testing criteria and evaluation. Finally, section 6 wraps up the main discussion and summarizes the contributions of our research endeavor.

## 2. Background and Related Work

In this section, we harness essential background information together with relevant research that holds some significance towards understanding the issues addressed in this paper.

A Security Operations Center (SOC) is neither a predefined entity nor a particular technological system that can be deployed to defend against specific security threats. Instead, a SOC is a static organizational structure underpinned by technological solutions to manage and enhance an organization's overall security posture. In the proposed framework, we plan to introduce and build adaptive SOC architecture irrespective of the organization's particular needs or to counter any potential threats. Adaptive SOC solutions will inherently address and tackle the multitude of varying organizational profiles [1].

A SIEM is a bundle of complex technologies that, together, can provide a comprehensive security layer in today's threat environment. Thus, the role of a SIEM component of a SOC cannot be overemphasized, and, as such, it must be carefully chosen. A modern open-source SIEM like Wazuh [15] has a centralized cross-platform architecture that allows it to control, handle and detect intrusions, threats, and behavioral anomalies on multiple platforms. Wazuh started as a fork of an OSSEC HIDS [40], adding more reliability and scalability to the project and providing an integration with the Elastic Stack as a log management solution and a RESTful API for easier task configuration and administration. Wazuh has two components: the Wazuh server component, which integrates closely with Elasticsearch and Kibana, while the Wazuh agent is capable of many security-related tasks such as log analysis, rootkit detection, listening port detection, and file integrity monitoring. For this project, we will utilize these capabilities to trigger alerts.

Furthermore, Wazuh provides the following mappings against regulatory compliance requirements:

- Payment Card Industry Data Security Standard (PCI DSS).
- NIST Special Publication 800-53 (NIST 800-53).
- General Data Protection Regulation (GDPR).
- Good Practice Guide 13 (GPG13).
- Health Insurance Portability and Accountability Act (HIPAA).
- Trust Services Criteria (TSC SOC2).

Although there is considerable amount of literature available to provide guidance on SOC implementation guidelines, there is little mention of any SOC automation for enabling horizontal scalability, expanding the security operations center toolchain, and supporting different risk profiles and threats.

The SIEM log management solution, and the systems adopted in our proposed framework abide by and readily comply with the guidelines and requirements in the resources as described below. This offers an excellent resource for developing SOC requirements:

- NIST Special Publication (SP) 800-92, Guide to Computer Security Log Management: provides guidelines on developing procedures for business logging and auditing [19].

- NIST Special Draft Publication (SP) 800-94 Revision 1, Guide to Intrusion Detection and Prevention Systems: recommendations are provided to plan, incorporate, configure, protect, track and manage IDPS technologies [35].

- NIST Special Publication (SP) 800-83 Revision 1, Guide to Malware Incident Prevention and Handling for Desktops and Laptops: provides recommendations for enhancing the security mechanisms of malware prevention by an organization and provides recommendations for improving the current incident management and response capabilities of an organization [38].

- NIST Special Publication (SP) 800-61 Revision 2, Computer Security Incident Handling Guide: provides guidelines to help companies build information security incident management, handling and response capabilities to resolve incidents quickly and effectively [29].

Slack [37], as a channel-based messaging platform, has been determined as one robust candidate to incorporate into our proposed framework. The primary value proposition of Slack is a tool for communications. This offers a communication mechanism that directly promotes workplace communication. It is evident, as a working piece of software, that Slack executes these roles incredibly well. As a result, Slack was well-positioned to address the surging demand on this requirement rather more effectively as offshore development teams in startups and working from home policies started to surface. Slack allows creating special channels for private communication between groups. These private channels can readily be used to organize big teams. Slack allows groups or teams to join a *Workspace* – a group of related channels via a URL or invitation issued by a team administrator.

Further, Slack offers users an application programming interface (API) to add apps and automate operations, such as sending automatic notifications based on human input and sending messages when a specific condition is met. The Slack API is one that offers great compatibility with various kinds of programs, platforms, and utilities. In our ongoing research endeavor we have sought to use Slack as a ticketing system for our SOC deployment. Further, our solution will automatically send any imminent alerts directly to Slack, where the SOC analysts can further investigate the nature of an alert and establish a ticket when necessary.

Ticket opening works by adding the Halp [11] which is a modern ticketing help desk to a Slack channel. The people of pertinence or the resident engineers at the customer organization are added to the channel, where they can view the open tickets and respond to the analysts during the time an issue is being addressed. In the sequel we leverage well-established results in certain pieces of the literature with the objective of better understanding SOCs, proposing suitable SOC frameworks, and assessing their immediate challenges.

In [6], the authors interview security analysts, in order to identify the technological and organizational limitations inside the SOC center. They propose a new workflow that ensures effective collaboration between the tiers in a SOC and the associated security incident correlations.In [17], the authors review existing, industry-accepted maturity models wherein the proposed SOC classification model complies with this approach. In [36], the authors define their SOC and demonstrate a method for assessment of any SOC, where they suggest a SOC framework commensurate with their conjecture. In [7], the author presents the initial model as developed by the WLCG SOC Working Group, for a minimally viable SOC, where he starts out with the design of different stages, elaborating upon the individual stages involved. In [3, 4], SOC experts have evaluated the use of Sonification as a form of human-computer interaction to improve upon the findings of the SOC research communities in anomaly detection, monitoring, and alerts discovery. In [2], the authors have noted the necessity regarding the existence of a SOC at educational institutions, by proposing threat-addressing courses, with current security tools in use, while proposing their own SOC model.

In [23], the authors propose a SOC architecture, together with its mission and key roles that act as the appropriate incident response mechanisms to recognize incidents of information security, resolve potential vulnerabilities, and restore compromised Internet of Things infrastructures. In [20], several methods of data retrieval are proposed to enable a data triage by retrieving the security analysts' pertinent historical data triage of operations involved. In [22] the authors conduct an interview that deliberate what organizations ought to do to reap full benefits from their investments in SOCs, where it further addresses the common mistakes security personnel often make when implementing a SOC.

In [27], the authors address the risks of cyber-attacks on businesses and the costs they tally upon the economy. The authors also suggest the adoption of SOCs to regularly and diligently track the vital and essential digital business services undertaken to protect and secure the best interest of the underlying business operations. They also suggest dividing a SOC into three significant categories: collecting information, running it through the appropriate analytics, in addition to monitoring and responding with countermeasures. Further, the paper addresses the role of each security personnel involved in administering the SOC.

In [24], , the authors give a brief overview of information security and then move on to the concept of SIEM solutions; deemed as central to the health of a functioning SOC. In [25], the authors discuss the risks of cyber-attacks and their financial impacts on the economy. Several SOC tools were brought to bear, along with best practices in applying processes and procedures.

In [10], , the authors introduce a user-centric machine learning system that leverages big data to classify and identify a risky user from different security logs collected, issue alerts, share relevant information, and offer analyst expertise all to determine user risks leveraging actual and real-life industrial data. In [18], , the authors elaborate upon some security threats stages towards SOC and suggest several solutions to safeguard the system, including Standard Operating Procedures (SOP) that make the SOC process and its associated actions more explicit. They also discuss regular training for employees on the company's functions, resources, policies, and standard rules in institutional operations. In [8], the authors propose an innovative intelligent framework referred to as Network Flow Forensics Framework (NF3). According to the authors, this framework is considered effective and accurate due to harnessed machine learning, network traffic analysis, and encrypted traffic identification. In [9], the authors introduce a novel intelligence-driven cognition-based computing SOC that is essentially based on progressive and entirely automated procedures. This is based on using two precise new and innovative artificial intelligence algorithms, wherein it harnesses the Lambda machine learning architecture that can process a combination of batch and streaming data simultaneously.

In [5],the SIAC architecture presented addresses an enterprise SIEM built on open-source technology that heavily leverages Wazuh's SIAC's main contribution, where it illustrates how open-source tools can be used to build a SOC with upscale information security capabilities. It further introduces the concept of automated SOC deployment via shell scripts and later reveals that it can be imported into configuration management/orchestration tools easily; however, it does not provide any actual implementation details.

It is important to note, nonetheless, that many of the provided scripts are either considered obsolete by today's operational frames or are not needed for Wazuh installation altogether. The following list offers a brief comparison between the two solutions:

- Our proposed solution, herein, is fully scalable at all levels including horizontal scalability with the ability to enhance the toolchain, add plugins, handle more agents, and support servers and clusters. In comparison, SIAC is scalable in terms of the ability to add Linux agents by only providing a shell script to install the agent with the needed tools to configure them.

- Our proposed solution offers the ability to create new alerts that can be mapped to the sought-after compliance requirements. In comparison, a SIAC entirely relies on already implemented mappings in Wazuh without providing any changes.

- From a holistic perspective, we have fully automated the proposed solution from A to Z. It is user-friendly, where the less-experienced users can leverage our proposed framework to configure any network regardless of its size. In the meantime, SIAC provides some simple scripts to install the required packages in case of any automation procedure requirements. However, the underlying codes are obsolete and require manual handling.

- In terms of adaptability, our solution is found to be fully adaptable in terms of risk profiles, network size, technological diversities, varying organizational requirements, and fluctuating organizational budgets. Meanwhile, SIAC can run in the cloud, on bare metal, or in a hybrid environment, that is supported by the open-source solution currently in use. However, SIAC does not seems to offer much in terms of adaptability.

- Regarding configurations, we have not paid any heed opposite particular unique configurations as it falls outside the realm of the immediate focus of our project. Nonetheless, many configurations are being considered towards achieving adaptability. SIAC, on the other hand, adds in a new plugin for better dashboard visualization (Kibana Network Plugin) together with a new package (Packetbeat) to enhance the logging processes and management side only.

- Regarding cloud support, it is readily seen that both solutions do, in fact, support cloud deployment. However, SIAC supports the cloud through the capability of the used open-source SIEM itself.

- Opposite the issue of modularity, we offer a fully flexible and automated solution in terms of architectural modularity. Any new module can be readily added to the deployed agents and the toolchain. In comparison, SIAC's capability of improving the toolchain is done by altering the provided scripts. However, since these are regarded as bash scripts, this method is again limited solely to Linux machines.

- We provide modifications to the existing solution whereby a new subsystem is added to the SIEM and provide new frameworks to enhance the overall capabilities of SOC. SIAC, on the other hand, offers nothing more than FOSS packages and configuration files with no significant modifications in the process.

## 3. Analysis and Design

The design of our proposed framework is intended to be as simple as possible, making the deployment of a SOC a straightforward matter. Yet, the design will leverage a powerful process to achieve what we refer to as an Adaptive SOC implementation. Automation of the underlying process is meant to cover the vital parts in a SOC deployment. Here, configuration of the monitored devices should entail fully automated processes. This encompasses agent installations on the client devices and configuration of network and security attributes in a manner that would allow the logs to continually update the central SIEM; a process that is normally rather tedious with repetitive task executions making it more error prone.

Furthermore, our framework design aims to reinforce an open-source-based SOC's operational and functional environment. This is done by fostering a plugins system that can interact with established applications, execute custom codes, and enhance the SIEM functionality.

1. Our SOC should be capable of the following:

- SIEM: provides the necessary security analytics including monitoring, and event correlation by allowing log collection, aggregation, parsing, storage, analysis, search, correlation, archiving, and disposal.

- Log management: helps ensure that records are stored in sufficient detail. The collection processes will handle the logs; it will gather data from different sensors and convert them into a unified standard format.

- File Integrity Monitoring (FIM): checks crucial files (e.g., operating system's files) to determine whether they have been tampered with or gotten corrupted.

- Incident Response: a system that facilitates, recognizes and responds to facilities, staff, protocols and communications to operate coherently in an emergency.

- Asset Discovery & Asset Management: keeps track of all active and inactive assets in order to continuously check the adherence of each asset to the overall security policy and to take appropriate action if the asset deviates from the policy.

- Threat Intelligence: provides evidence-based information of a current or emerging threat, including context, mechanisms, indicators, consequences, and actionable recommendations.

- Rich Analytical Dashboards & Data Visualization: visually presents the complex data sets and outcomes of the analysis done by other SOC components.

- Real-time and Configurable Alerts & Reporting tools: provide real-time notifications of detected anomalous events, with information regarding what, where and when events happened.

2. Regulatory compliance: the conformity of an entity to rules, regulations, standards, and requirements applicable to its business processes. Some compliance regulations are explicitly intended to ensure the security of records and protection of data. Bad compliance practices for data breaches can harm customer satisfaction and adversely affect a company's bottom line. As data breaches continue to rise in frequency, customers become more inclined to investing more confidence in enterprises that strictly obey regulatory compliance mandates intended to safeguard and protect personal data. Data privacy-specific regulatory compliance mandates, such as General Data Protection Regulations (GDPR), have become more prominent as the processing of users' sensitive data by corporations is becoming more scrutinized. Given that a SOC is a requirement that must be met by hospitals, banks, and other governmental organizations, the proposed SOC implementation would readily provide vendors with features that reflect conformity to regulations. This is owed to the fact that clarity in compliance to procedures often gives clients more confidence in the underlying business processes. Besides, compliance with legislation serves as one of the elements that ensures security of the data involved.

3. The Ticketing System: It is expected that a SOC team will monitor potential incidents identified by tools or reported by charged personnel. To guarantee that events are properly handled, a case must be established, allocated, and monitored before completion. Both tools and personnel involved should help fulfill the processes involved, provided that the proper equipment, jurisdiction, and integration of incident response and case management procedures are in place. A crucial factor to remember is that remediation for some incidents might require resources outside the reach of SOC analysts' reach. Therefore, the effectiveness of case management is manifested in its ability to assign these responsibilities to their respective lines of authority. Thus, the ticketing system, which is used to track events through its incidents history as well as a contact point between the impacted segment and the SOC, sits at the forefront of the a SOC incident handling process.

4. Automation: One of this framework's main goals is to achieve a fully automated SOC deployment, better known as a plug-and-play property. This automation feature aims to aid the SOC to meet the challenges that emerge from many determined and repeated processes. Therefore, the deployment and configuration of the monitored devices should be fully automated offering high-velocity tasks. This requires that we establish the first pillar of the Adaptive SOC tooling, which shall henceforth be referred to as the "*Autoconfig Tool*".

   The proposed framework should be able to fully automate the following processes:

   (a) Configuring the entire SOC stack.
   (b) Deploying agents on monitored devices.
   (c) Configuring monitored devices for agent-less monitoring.
   (d) Deploying and integrating the SIEM with a ticketing system.
   (e) Modifying the configuration of the SOC components at any time.

   With this property, becoming active, the SOC will achieve:

   - Configuration management and full SOC deployment in minutes.
   - Help bridge the cyber skills gap through automation that lessens the burden on humans.

5. Contriving Horizontal Scalability & Modular Architecture: The SOC's security space should foster the ability to be expanded with new tools. Furthermore, the SOC should possess the versatility to increase its capacity and include more servers on clusters as a single logical unit. Here, improving the capacity, as well as improving the security space, should be very seamless processes without any added complexities.

6. Fostering Adaptability: Exposure to the Internet is presently a fact of life inherently posing numerous threats upon system integrity. SIEM solutions can provide much information to help the end user quickly take actions against security threats and accurately identify the attacks, all from a single console. Yet that may not be enough; the SOC should enhance its maturity and ability to mitigate incidents. In order to have a proactive defensive strategy against the different risk profiles of different organizations, or even against the unexpected threats, the SIEM must be able to work together with another module; a module that can provide a broader range of contextual information — about identities, users, device types, privilege levels, conditions of networks and events –a functionality usually embodied in the SIEM.

This requirement originated from the problem that the current open source based SOC implementations lacks the ability to be strengthened on the fly, and since their development is usually a slow process, complicated with the unavailability of receiving frequent updates (as the commercial ones), the need for a tool that can promote the capabilities of the security solutions, without needing to upgrade the SOC, is a must.

The Adaptability enforced by what we refer to as the "Plugins System" behavior will stand as the second pillar for an Adaptive SOC tooling. It enables every SOC manager to adapt to different risk profiles and to deploy new functionalities that can assess security threats and take immediate remediation actions on the monitored devices in the form of plugins, in real-time. These plugins should be easy to develop, easy to expand, and easy to maintain. Furthermore, these plugins should be easy to share; to enable a security ecosystem of collaborative defense that extends available security solutions with easy to install plugins. Such transparent and collaborative strategy spurs the development of new plugins that further enrich the cyber defense world.

With this property on hand, the SOC ultimately becomes:

- Adaptable to different risk profiles and varying organizational needs.

- Amenable to deliver deep visibility and actionable insights into the complex threat landscape.

- Versatile to increased efficiency and performance for an SIEM solution with ready-to-install plugins.

- Amenable to a sharable experience in code development, exchange of best practices, gain insights and learn from others.

Before choosing a SIEM to satisfy our design requirements, we have conducted a thorough comparison between the available open source SIEMS. Our choices were broken down to:

- Wazuh

- Open-Source Security Information and Management (OS-SIM)

While both SIEM solutions above are very powerful, OS-SIM has clear disadvantages. This is attributed to lack of a centralized log management system. However, the definitive factor is, unlike Wazuh, which functions in a Manager-Worker fashion, OSSIM works in a single server architecture. Having single-server architecture deployment readily implies all computations will take place on a single node, requiring a very high-end machine to accommodate the burden; worse yet, it renders the network in a severely congested state.

The subsections in the sequel outline the different design strategy and the reasoning for each of our these choices.

### 3.1.1. Pull vs. Push Automation

The managed devices can receive their configuration parameters in two ways: the first being the Push model, which is the one we adopted in our research, with the latter being the Pull model, where the contrast is depicted in Fig. 1.

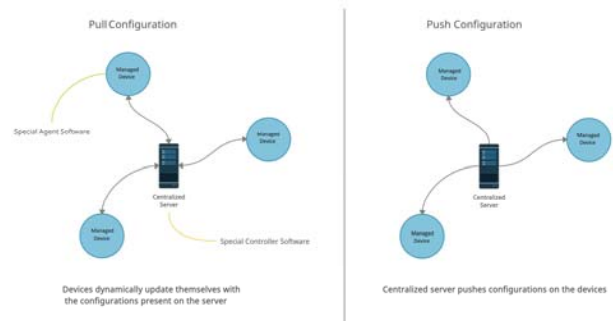

**Figure 1:** Push vs. Pull Configuration Management.

The push model requires nothing more than having the configuration tool and the SSH credentials available for the managed devices; it simply will log on to the managed devices, do the configuration, and the process is complete. Meanwhile, the Pull model requires far more than that in the Push model. Here, the managed devices would need to have a special agent that can communicate with the active manager, which means one level less of automation. This is because the managed devices will need to have the agent deployed before even starting the automated configuration process. Other disadvantages accruing from the use of the Push Model also include its difficulty to use, its slower performance, and its requirement for an SSL certificates management handler.

### 3.1.2. SIEM Integration vs. Host Plugins

Recent entrants to the SIEM landscape are adding technology that offers unnecessarily higher levels of sophistication for analytical use cases. These technologies can extend the SIEM functionality in two ways. The first uses an integration module while allowing the SIEM and other security applications to have third-party applications running on top of them or exchanging data with them via APIs.

However, the second approach is designed to deliver these technologies and applications and make them run locally on the monitored devices or hosts through the SIEM. This approach will allow the SIEM to have a broader view of what is happening on these devices. As such, it will continue to enhance the investigative capabilities and, consequently, will introduce a form of plugins for response actions. This is a powerful feature as it reduces the time to respond to security events in stopping attacks and protecting critical data rather more efficiently. From our own perspective, the Adaptable SOC will need to have both features on board. On this note, while the current SOC market is geared towards SIEM integration and empowering the SIEM with more sophisticated tools that pop everywhere, the second approach has been significantly less popular.

| Automation | Push | Pull |
|---|---|---|
|  | No software is needed to manage clients, simpler to use, and scripts written using YAML. | Needs special software on both the manager and managed devices, uses a unique syntax (scripting language) and has needless complexity. |
| SIEM | Wazuh | OSSIM |
|  | Under active development, well documented, very powerful and has scalable architecture. | No log management, no frequent updates, no manual or documentation and has Single-Server architecture. |
| Adaptability | Plugins | Integrations |
|  | Plugins can work as a separate system and can trigger the SIEM solution. Plugins are used to add new features inside the SIEM solutions to help the SOC in general and add new security tools as a means of threat detection and prevention based on the active response. | Integration triggered by the SIEM solution itself. Integrations can enhance the SIEM itself by adding a specific job into the SIEM solution without the ability to add new features or change the SIEM functionality (e.g., scan a file or send alerts into and outside the SIEM). |

**Table 1**
Different Design Approaches with Advantages and Disadvantages.

Table 1 summarizes the different design approaches citing advantages and disadvantages for each approach.

To satisfy the automation requirement, Fig. 2 shows the design details for the *Autoconfig tool* that is responsible for the Plug and Play SOC deployment and the horizontal scalability of the SOC. The tool will have a list of IPs of the monitored devices and the SOC servers, SSH logins, along with their corresponding roles as input; these roles include:

- The SOC component, e.g., SIEM server, Log management server.
- The network device operating system, e.g., IOS, JunOS.
- The endpoint device operating system, e.g., Linux, Windows.
- The security device.

The tool will then forward every IP address to the appropriate configuration module, which will take care of configuring the device either by setting up Syslog, deploying an agent, or configuring the SOC components and connecting them together.

As shown in Fig. 2, the tool will receive the required SOC topology as input. The SOC topology is a text file with a particular format to declare the devices IP addresses and their roles in the SOC; i.e., agent IP, SIEM IP, and so forth. The tool will have a parser that will read the topology file and parse the fields to choose the proper deployment module for this role, which takes place using Ansible [32] – renowned as a powerful push configuration tool. The result will be manifested in deploying and configuring the different SOC components and, hence, culminating into a fully operational SOC just via the sheer implementation of this particular tool. To satisfy the modular architecture and horizontal scalability requirements, the tool does not only support the initial SOC deployment, but also supports the capability of adding more tools to the SOC stack, deploying more agents and changing the configuration of existing deployments, as illustrated in Fig. 3.

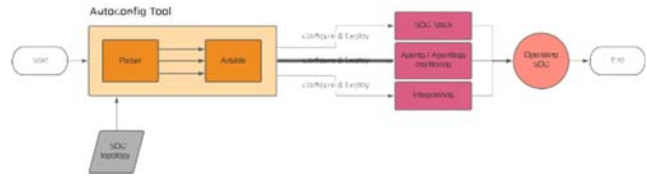

**Figure 2:** Autoconfig Tool Design.

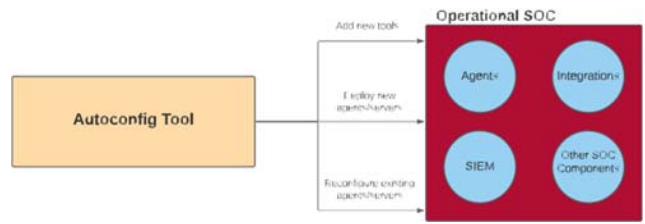

**Figure 3:** Horizontal Scalability with *Autoconfig Tool*.

This allows the SOC manager to deploy a new set of agents on-the-fly, add new monitoring tools to already existing functioning agents, add new integrations with SIEM, and expand the SOC clusters to adapt to the needs. The tool also follows and adheres to the following best practices:

- Safeguarding Sensitive Data in Encrypted Files: The *Autoconfig Tool* will need to use sensitive data, and the login credentials to configure the monitored devices in a seamless manner. Such sensitive data will be stored on the local file system for safeguarding purposes; these will be encrypted and require a passphrase to get them decrypted.

- Using SSH Keys in lieu of Passwords: In our proposed design, we abide by the best practice of using the Secure Shell (SSH), which implies an SSH key authentication mechanism instead of using passwords. Here, SSH key authentication is much more robust, and when sensitive data is transmitted across the network, security is of paramount importance.

The Programmable Plugin-based Intrusion Detection and Prevention System (PPIDPS) is designed specifically for security engineers. It offers the ability to design any threat detection mechanism. Also, it provides a prevention capability by triggering an active response technique. Furthermore, it lends itself well to programmability since it becomes the security engineer's responsibility to write the detection code and design the specific detection mechanism. Moreover, it is plugin-based, so security engineers worldwide can exchange these plugins and leverage them against different threats.

This system will act as a host-based intrusion detection system and a host-based/network-based intrusion prevention system (Depends on the programming) simultaneously. With this, it will run on all monitored devices with the purpose of adding the ability of enhancing the SIEM functionality and executing security functions directly on the monitored device themselves. The system will enable the SOC manager to execute custom code written in a scripting language or health check scripts as a form of a plugin on the monitored devices or a subset thereof and check the results in real-time. The results will be returned in logs through a secure channel and will then be fed to the dashboards. The design schematic shown in Fig. 4 presents code reuse with code sharing in mind; we believe, with this specific design strategy on hand, SOC engineers and analysts can readily share their plugins on threat intelligence or other platforms with great ease.

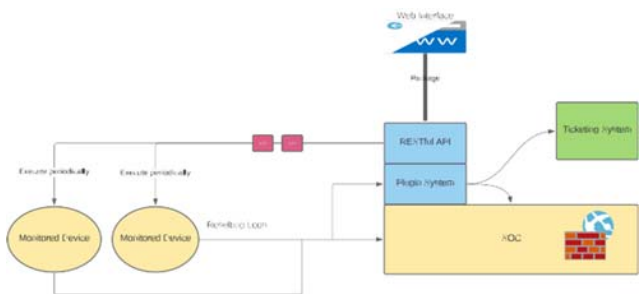

**Figure 4:** Programmable Plugin-based Intrusion Detection and Prevention System Design.

Our Plugin System uses a RESTful API that runs on the Wazuh manager. This RESTful API will use an encrypted channel to push data to the monitored devices. The data can be some custom scripts written in a platform-independent language (e.g., Python3), or health check codes, or simply a series of commands. The plugins can be fed into the Plugin

System using the RESTful API. A web interface is provided to allow the users to configure, edit, and push new plugins into the system using a simple User Interface (UI). The plugins are pushed as packages into the system; these packages include:

- The script that is going to run on the agent machine and report for possible threats.

- The needed decoders and rulesets for the logs produced by the plugin.

- The needed scripts to execute a proper incident handling mechanism to resolve the threat.

The SOC manager would enable those plugins to run on the deliberated devices. Enabling the plugin will signal the agents on the monitored devices to download the plugin and execute it during every configurable period. The needed decoders and rulesets to produce alerts are pushed into the SIEM once the plugin is enabled. The scripts will check on the monitored devices or, in the case of a new threat appearance, run custom scripts to make these devices threat immune. The agent on the monitored devices will keep running the scripts, and upon discovering a threat, the agent will send the alert data in the form of Syslog to the SIEM. The SIEM will use the associated decoders and rulesets to create proper alerts and push them on the dashboards to give the security analysts a holistic picture of available information. The Plugin System's triggered alert will open a ticket using the Ticketing System, with the highest priority – "critical". The Plugin System will also take any required incident handling measures to prevent the threat from migrating or moving to other parts of the system; for instance, upon adding new configurations into the firewall.

Since the needed decoders and rulesets are bundled together with the plugin scripts, this renders the plugins fully programmable. Any part of the plugin can be altered to satisfy the different needs and different risk profiles of different organizations. For example, the decoders and rulesets can be modified to trigger different alerts that are more suited for a particular organization. The alert levels can be tweaked to fulfill the needs of the risk profile of a specific organization; better yet, the alerts can be mapped onto regulatory compliance requirements that a specific organization has to adhere to.

### 3.1. Engineering Standards

Our framework follows the following Engineering Standards in order to provide a stable continually evolving foundation that enables entire industries to develop and thrive.
List of these Engineering standards include:

1. Log Format: The format of logs for Linux systems and network devices sent by the underlying tools in the proposed framework will follow the message logging standard, Syslog as defined in [13]. Since Syslog provides no data encryption by default, the Syslog implementation as defined in [26] uses TLS to encrypt the Syslog traffic. For the Windows operating system, Eventlog is used.

2. Transmission Control Protocol (TCP): The exchanged data in the communication medium will be handled using the TCP protocol as presented in [16].

3. Secure Shell: Secure Shell (SSH) is a communication protocol for remote login that is handled securely over insecure networks, as defined in [42]. SSH will be used for configuration management automation purposes.

4. Encrypted Channels: All the catered for tools will have to exchange information across the network, be it in the parameter configuration and the custom codes to be executed on the monitored devices, including other private information. These exchanged data packets are considered sensitive and therefore must be maintained confidential; hence, we add SSL encryption to all our tools communications.

# 4. Implementation

In this Section, we reveal the implementation details of our framework; towards the end, we present a use case that shows how the entire system will function, as one entity, to yield the desired outcomes.

## 4.1. The *Autoconfig Tool* Implementation
### 4.1.1. Ansible

Ansible [32] is a push model configuration tool, that has recently prevailed and became one of the pillars of the DevOps world. It can configure systems, deploy software, and orchestrate IT tasks. Ansible was built with simplicity at its core, with its operation not needing any client or agent to be installed on the target machines. Unlike other tools, including Chef [30] or Puppet [21], it only requires an SSH connection for it to execute its allocated tasks.

### 4.1.2. Ansible Architecture

Ansible uses SSH for communication with the target machines, where it can run commands directly on them, or receive the information that they send. Ansible has a very simple architecture, which consists of three main components:

1. **Inventories**: These are used to save the host information as text. The information might include the IP addresses, the SSH keys, the username, passwords, URLs, checksums, etc. The hosts inside the inventory can be divided into groups to reflect corresponding configuration needs.

2. **Playbooks**: These are units of scripts that will define what configuration or any piece thereof will run on the target machines. These playbooks are defined in YAML format and extensively make use of the Modules.

3. **Modules**: These are standalone scripts, written in Python, which can be used by the Ansible-Playbooks. These modules can control services or daemons that run on different operating systems, system resources, filesystem content, or for installing packages, etc. and can execute raw commands. Ansible comes with a large set of ready modules that can be extended by adding one's own modules.

Furthermore, Ansible can execute raw commands using the shell or command line. Fig. 5 clarifies the interaction between various Ansible components.

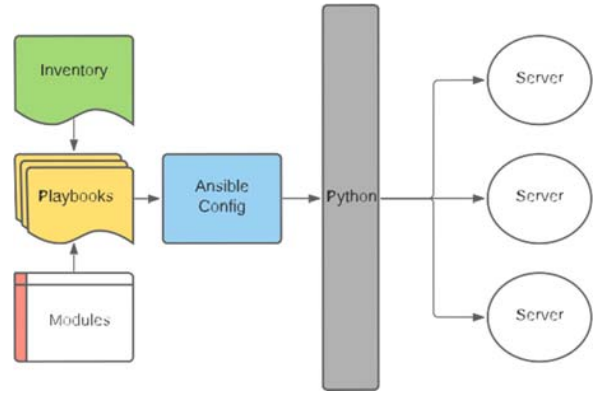

**Figure 5:** Ansible Architecture.

### 4.1.3. Ansible Vault

Ansible-Vault is Ansible's mechanism of providing encryption and security to the sensitive data across inventories, in a user-friendly manner [33]. Vault will ensure file-level encryption of sensitive files and require the same passphrase for encrypting as well as decrypting the files. Ansible-Vault will be used to ensure security of SSH keys in the implemented tools.

### 4.1.4. Jinja2 Templating Engine

Ansible uses Jinja2 [28] templating engine to replace the placeholders with appropriate values. This, offers greater flexibility in the written playbooks. For example {{ wazuh_ip }} will be replaced by the value of the wazuh_ip variable. Furthermore, this engine will bolster the ability of using conditional statements (*e.g., if statements*) and flow control statements (*e.g., for loops*) [34].

## 4.2. The Automation Engine

Given Ansible's great potential and automation capabilities, as described in [41],it is readily evident that Ansible can reduce the time for configuration and provide easier network devices maintenance. This is on top of the added capabilities afforded to legacy network elements to operate in a similar manner commensurate with upscale network devices, with Software-defined networking (SDN), performance. The *Autoconfig tool*, under the proposed architecture will heavily leverage Ansible as an Automation Engine. In so doing, one would be able to automate every phase of Wazuh's deployment, execute endpoint configurations, and do necessary in-tegrations with external APIs. Hence, the following set of Ansible-Playbooks will be provided:

1. Wazuh all-in-one deployment playbook
2. Wazuh distributed deployment (Wazuh clusters) playbook
3. Wazuh Windows agent deployment playbook
4. Wazuh Linux agent deployment playbook
5. CiscoIOS configuration playbook
6. JunOS configuration playbook
7. Wazuh integration with a Ticketing System (Slack) playbook
8. Wazuh integration with VirusTotal playbook

Each playbook will harness different modules and configuration parameters that are mostly suited for the task on hand. These playbooks will decouple the deployment process from the lower details of how the agents or the Syslog configuration works for different platforms. At any point in time, be it at the initial deployment of the SOC or when expanding a SOC with new monitored devices, the SOC manager will only need to gather their IPs and SSH keys and get the job done in a matter of seconds. The same procedure applies if the SOC manager decided to extend the SOC with a Ticketing System integration or another API. Here, the SOC manager will not have to go through the cumbersome process of remembering lengthy details of how a particular task would be done. This will also spare the extra time normally needed in searching within the docs. In summary, all that needs to be done would be to run the relevant playbook and get things done, as per the characterizing plug-and-play property. Fig. 6 illustrates the entire process in detail.

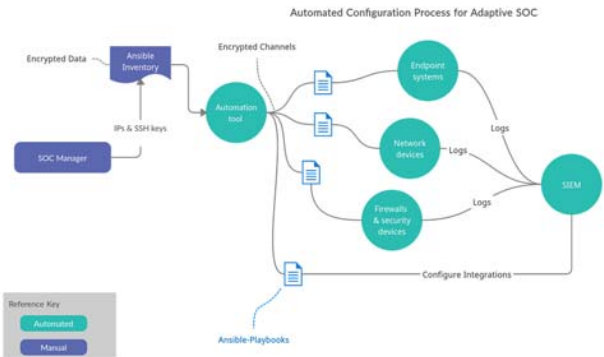

**Figure 6:** Automated Configuration Process.

In executing the processes seamlessly, the different roles of an Automation Engine are split into different directories to evade any repetitive activities and to aid into the process of integrating the engine. The file structure of the Automation Engine is shown in Fig. 7:

- **playbook directory** contains all the playbooks that covers the different cases of deployment; this includes the agents, the Wazuh and Elasticsearch clusters, etc.

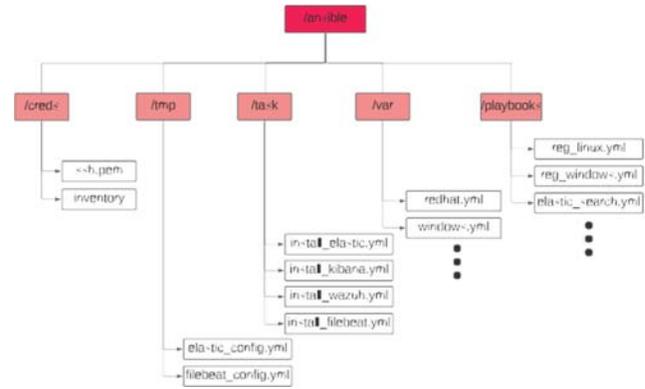

**Figure 7:** The Automation Engine File Structure.

- **vars directory** contains the needed variables that will be replaced in the appropriate playbook, e.g., IP addresses.

- **tasks directory** contains common tasks to be imported by the playbooks, to evade needless repetitive executions of the same codes, e.g., installation of the Elasticsearch package.

- **tmp directory** contains configuration files that will be pushed into the deployed machines to adjust their configs to a client's needs, e.g., Filebeat configurations.

- **creds directory** contains the credentials that will be used by the playbooks to gain access to the machines; these files will be encrypted using Ansible Vault, e.g., SSH keys.

If the SOC manager wants to deploy an Elasticsearch node, he or she would simply run the relevant playbook. In this case, the "elastic_search.yml" playbook. The playbook will import the needed variables and tasks from /vars and /tasks, respectively, and will use the relevant credentials from the /cred directory to login to the specified machine in the playbook and run the tasks; this, in turn, will deploy Elasticsearch. In the process, the defaults configuration of Elasticsearch will be replaced by the appropriate ones in the /tmp directory.

While users can install and deploy any of the components by running its playbook at any time, the users may, at some point, want to make a huge deployment of different components, e.g., during the initial deployment phase. Therefore, we provide an interactive utility which the user can run prior to making the deployments called *formatter.py*. This utility works interactively with the user, while helping abstract the tool and serves as an interface. It will ask the user for the number of Wazuh servers needed and the number of Elasticsearch clusters to be deployed, and the integrations that are desired for adding to Wazuh. Finally, it will ask the user for the number of agents of a specific type that will be deployed.

For each type of deployment, the tool would also ask for the machines' IP addresses, along with the path for the SSH keys and the SSH username. Alternately, *formatter.py* could run with -a or -agents options for the sole purpose of acquiring input for the agents.

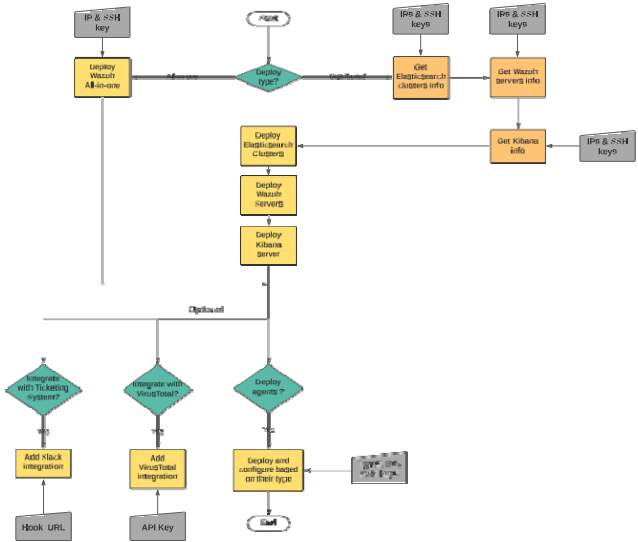

**Figure 8:** Autoconfig Tool Workflow.

The result from this utility is a text file (by default called *topology.txt*) which contains the entered information in the format shown below:

IP: SSH Key Path: Device Type: SSH User

The *Device Type* defines the role of the machine. It can be one of the following:

1. **Linux** (agent), with the ability to distinguish between the different distributions.
2. **Windows** (agent)
3. **Cisco**
4. **Juniper**
5. **Elastic** (server)
6. **Kibana** (server)
7. **Wazuh** (server)

The resulting text file from running the *formatter.py* will be fed into the deployment script – *deploy.py*, which will invoke the actual deployment process. Depending on the *Device Type*, the script will forward the information to the appropriate Ansible-Playbook, which, in turn, will take care of both the configuration and deployment processes. Fig. 8 illustrates the two scripts in action.

## 4.3. PPIDPS Implementation

The PPIDPS is implemented by four major components:

1. **The Plugins**: These reside on the Wazuh manager. They contain the actual threat detection capabilities and are used by the agents.
2. **The SOCreative API**: A RESTful API that runs on the Wazuh manager. It is used to interact with the plugins system through HTTP requests. It can be used to push, pull or update plugins. "SOCreative" is a name that has been introduced, by the authors, for the API.

3. **The SOCreative Web Interface**: This is a web interface that will be used by SOC engineers. It is used to interact with the API itself and manage the plugins system with ease.
4. **The Agent Daemon**: This is a daemon that runs locally on the agents. Its purpose is to keep track of which plugins to download from the API and to execute those plugins on regular basis and report the results to the Wazuh manager.

## 4.4. Plugins

All the internal codes to build this plugin system are implemented defensively, i.e., the codes will exit gracefully if they fail during execution and print where they failed on the screen.

### 4.4.1. Structures

The plugins system directory will have the needed files and sub directories that will be used by the plugins system:

- The plugins implemented by the security engineers should be located at `/var/ossec/plugins` directory, each of which will be identified by a unique ID.

- Every agent that is intended to receive a plugin will have a JSON file named pursuant with its ID and located in the `/var/ossec/etc/shared/default/plugins` directory. This directory is shared between the manager and the agents using the Wazuh daemons.

- The new decoders will be installed and located in the `/var/ossec/etc/decoders/local_decoder.xml` file.

- The new ruleset will be installed in the `/var/ossec/etc/rules/local_rules.xml` file.

- The scripts that are intended to be executed for active response purposes will be located at the `/var/ossec/activeresponse/plugins` directory, each of which will be named by the plugin's ID they relate to.

The plugins will be pushed to the */var/ossec/plugins* directory. Every plugin will be named after its assigned id. The plugins directory structure is shown in Fig 9. The plugin can be pulled by the agent using a GET request to the RESTful API, where the agent will receive a minimal size of the plugin. This includes only the *<plugin's id>/script.py* and the *<plugin's id>/metadata.json*. Other files are only needed by the Wazuh servers. The *decoders.xml* and *rules.xml* will contain the needed decoders and rules templates by the Wazuh server to interpret the received logs by the plugin and trigger the appropriate alerts. The template in Fig. 10 will be followed by the decoders in the *decoders.xml* file. In addition, the template in Fig. 11 will be used in the *alerts.xml* file.

The *script.py* is the script that will be executed locally on the agents. The script will contain the required functionality to do the health check or identify an attack. The script will also contain the required code to provide the ability to achieve remediation and apply any needed countermeasures.

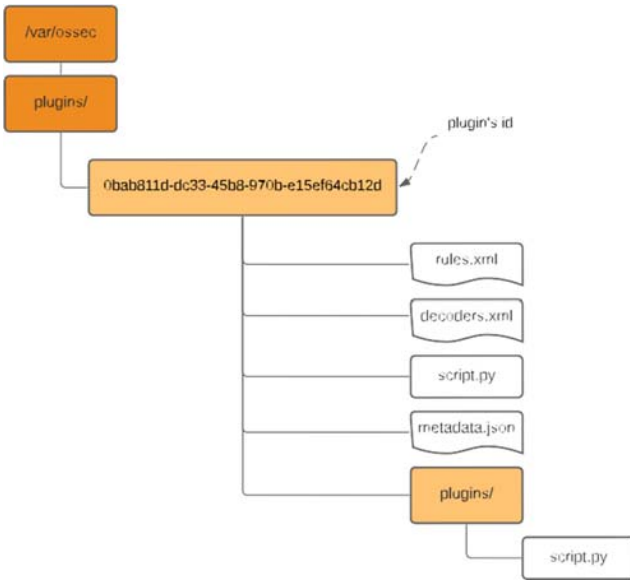

**Figure 9:** Plugins Directory Hierarchy.

```
1  <decoder  name="DecoderNameForThePlugin">
2    <prematch>\.∗SOC_NES: (\.+)</prematch>
3  </decoder>
4  <decoder  name="DecoderNameForThePlugin">
5    <parent>DecoderNameForThePlugin</parent>
6    <regex>(TheNameOfThePlugin): (Value_1) (Value_2)
7  (Value_3)</regex>
8    <order>pluginName, val1, val2, val3</order>
9  </decoder>
```

**Figure 10:** PPIDPS Decoder Template.

```
1   <rule id="ruleID_For_DecoderX"  level=
2   "AlertLevelToBeTriggered">
3     <decoded_as>DecoderNameForThePlugin</decoded_as>
4     <description>TheNameOfThePlugin has been triggered
5     </description>
6     <field  name="pluginName">TheNameOfThePlugin</field>
7     <field  name="val1">The Value Of val1 variable</field>
8     <field  name="val2">The Value Of val2 variable</field>
9     <field  name="val3">The Value Of val3 variable</field>
10    <group>TheNameOfThePlugin</group>
11  </rule>
```

**Figure 11:** PPIDPS Rule Template.

The *metadata.json* will contain metadata about the plugin and will have a structure as the one shown in Fig. 12.

The *id*, *name*, and *description* represent the unique id of the plugin, the given name to the plugin and a short description about the plugin's functionality, respectively. The version reveals the version of the plugin and must be in incremental order. The *enabled* represents the current state of

```
1  { "id": "0bab811d−dc33−45b8−970b−e15ef64cb12d",
2      "name": "Plugin Name",
3      "description": "Plugin Description",
4      "version": "0.0.1",
5      "enabled":  false,
6      "script":  {"interval":  60},
7      "agents":  ["001", "002"]
8  }
```

**Figure 12:** PPIDPS METADATA Template.

the plugin; it is either enabled and, hence, executed by the agents or is not enabled, altogether. The *interval* represents the period frequency in seconds in which the script will be re-executed. The *agents* represent a list of agents' ids that will receive this plugin and have it activated.

### 4.5. Active Response Mechanism

We have implemented our own active response system with the purpose of gaining more fine grain control over the monitored devices and react accordingly in the case of a threat. In other words, to enable an appropriate incident handling mechanism.

The system can take effect on both the agent device and on the Wazuh manager itself. It takes place on the agent device by including it in the main plugin script; i.e., *script.py*, where the script will have all the needed logic to take the proper actions to react and do the needed remediation actions once it detects a threat. It will then send a request to the Wazuh manager along with a list of arguments that might be needed – to start executing the active response on the Wazuh manager locally.

The active-response directory (inside the plugins directory) will contain the required scripts that need to take place on the Wazuh manager (e.g., adding firewall rules or quarantining the device) and will be applied by running the *active-response/script.py* script. It will have the structure shown in Fig. 13.

where the active-response scripts will be named after the plugin's id that they are associated with.

### 4.6. SOCreative API

We have used a high-performance Python framework, called FastAPI [31], to run the RESTful API on the Wazuh manager machine on port 55002. Through this API, the SOC team will be able to directly export/import the plugins from/to Wazuh. The API will have the following endpoints:

- **/plugins/** - GET: retrieve a list of the enabled plugins.

- **/plugins/** - POST: import a zipped plugin.

- **/plugins/{plugin_id}** - DELETE: disable and then remove the plugin's directory and files.

- **/plugins/template-plugin.zip** - GET: retrieve the template to create plugins, this becomes handy to use as a guide to create new plugins easily.

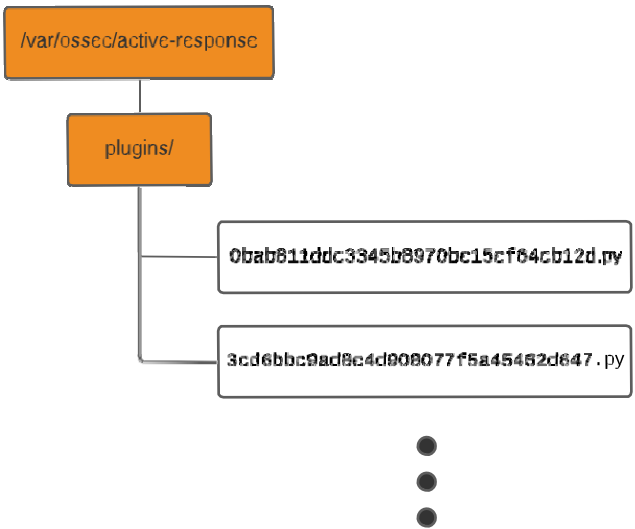

**Figure 13:** Plugins Active Response Hierarchy.

- **/plugins/{plugin_id}.zip?size={size}** - GET:

  - plugin_id: the id of the plugin to be exported.

  - size: the size of the exported plugin can be full or minimal; the minimal size includes only the *script.py* and the *metadata.json*, which are intended for the agents since the daemon that runs on the agent device requires just the availability of these two files.

- **/plugins/{plugin_id}.json** - GET, POST: retrieve or update the plugin metadata e.g. enabling the plugin, changing the agents list or version.

- **/plugins/{plugin_id}/ar** - POST: retrieve the arguments needed to run the active response scripts.

### 4.6.1. Enabling Plugins

Whenever a new plugin is enabled (using the `/plugins/{plugin_id}.json` endpoint) or when the plugins are declared as enabled once they get imported unto the system (using the `/plugins/` endpoint), the following actions would automatically take place:

1. The local decoders and local rules of the Wazuh system, `/var/ossec/etc/decoders/local_decoder.xml` and `/var/ossec/etc/rules/local_rules.xml`, will be appended by *decoders.xml* and *rules.xml*, respectively, things that are provided by the enabled plugin.

2. The *active-response/script.py* file will be moved to the `/var/ossec/active-response/plugins/` directory and will be named *{pluginID}.py*.

3. Wazuh manager will restart the wazuh-manager service to apply the changes.

4. For every agent defined in the agents' list at *metadata.json*, a new *{ID}.json* file will be pushed into the shared directory, `/var/ossec/etc/shared/default/plugins`, where {ID} represents the id of the intended agent where it will ultimately be used. Here, every added agent will be notified about the update. The directory will have the structure shown in Fig. 14.

Every *{ID}.json* file will have all the enabled plugins for this agent. The structure of the file is illustrated in Fig. 15, where the "ID" represents the identifier of the plugin to use. However, having the "version" variable available which caters for different versions of the same plugin for different use cases.

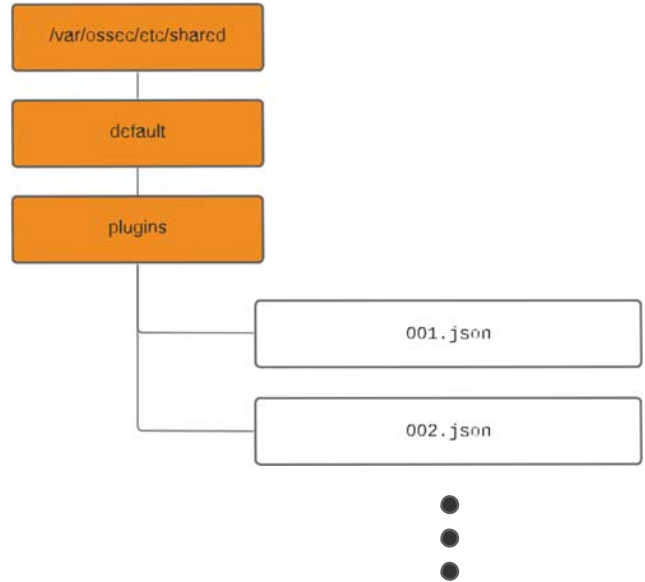

**Figure 14:** The Shared Directory Structure and The Distribution of The Flag Files.

```
[ { "id": "0bab811ddc3345b8970be15ef64cb12d",
    "version": "0.0.1"},
  {"id": "3cd6bbc9ad8e4d908077f5a45462d647",
    "version": "0.1.0"}
]
```

**Figure 15:** PPIDPS Flag File.

### 4.6.2. SOCreative Kibana Plugin

To make the interaction with the API more convenient, we have embedded a web interface that is intended to interconnect with the API natively into Kibana's web interface. SOCreative Kibana Plugin is compatible with Kibana's version 7.9.1. Using this interface, SOC engineers will be able to import new plugins, remove those unneeded, enable and disable existing ones, while updating others. The main interface is shown in the Fig. 16. Meanwhile, by clicking on the "Plugins" module, the window in Fig. 17 will appear.

As shown in Fig. 17, and by using **(1)**, the SOC team can view existing plugins. The "Enable" button can be used to enable or disable the corresponding plugin. Whenever a plugin is disabled, the following actions occur:

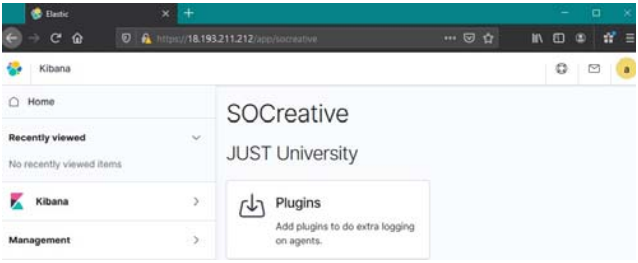

**Figure 16:** PPIDPS Main Interface.

1. The API will remove the *decoders.xml* and *rules.xml* contents from */var/ossec/etc/decoders/local_decoder.xml* and */var/ossec/etc/rules/local_rules.xml* files at the manager.

2. Restart Wazuh manager service to apply changes.

3. Remove the disabled plugin id from the agents JSON files at the shared directory */var/ossec/etc/shared/default/plugins*.

4. The agents will be notified of the change and disable the plugin.

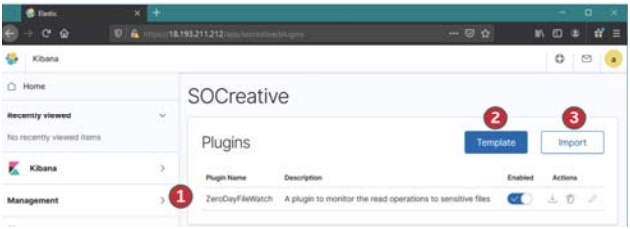

**Figure 17:** PPIDPS Plugins Interface.

Furthermore, the SOC team can download the plugin and view it locally or remove the plugin entirely (which will trigger the same actions as disabling a plugin plus removing the plugins files) or just simply edit the plugin. Using the edit module will readily open the tab as shown in Fig. 18. However, by using this tab, SOC engineers can change the *metadata.json* file contents directly. The name, description, version, the agents' list, and the code interval can be changed altogether. Here, whenever the agents list is updated, the following mathematical operations set will be applied:

$$R = O - (N \cap O)$$

$$A = O - N$$

Where *R* is the set of agents which will have the plugin removed, *A* is the set of agents that will have the plugin installed, *O* is the old agents set and *N* is the new agents set.

Then the IDs defined in set A will be added to */plugins-system/default/plugins* with the plugin ID and version and automatically distributed to the agents by the daemons. The IDs defined in set R will be removed from the directory and automatically removed by the daemons from the agents.

Lastly, **(2)** is used to download the template plugin locally, and **(3)** is, then, used to import a new plugin as a zipped archive. Figure 19 shows the import module just referenced.

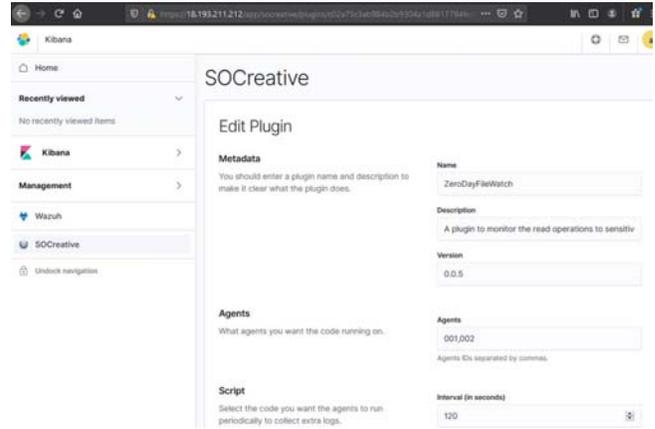

**Figure 18:** PPIDPS Plugins Editing Interface.

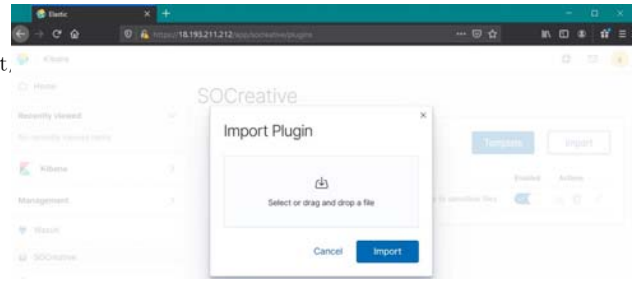

**Figure 19:** PPIDPS Plugins Importing Interface.

### 4.7. Agent Daemon

We have implemented a new daemon that will run on the agents, using Python 3, called *agentd*. This daemon will be responsible for pulling the scripts, executing them, and sending the resulting logs to the manager.

#### 4.7.1. Structure

The daemon will create the following directories and subdirectories to use them for its operations:

1. */shared/plugins*: This directory will receive the new plugin IDs, and their version numbers, which are intended to be used by the agent from the daemons. The data will be contained in a JSON file named by the ID of the agent. This file is a clone of the *shared* directory at the manager.

2. */plugin_download*: This directory will be filled by the activated plugins for the agent. Every plugin's directory will be named pursuant with its plugin ID.

These directories will be automatically created by the daemon upon its initiation. For Windows, these will be created under *C:\Program Files (x86)\ossec-agent*, and for Linux they will be created under */var/ossec/etc/*.

#### 4.7.2. Installation

To install or delete the *agentd* on the agents, the following command is used:

```
user@nes# ./agentd.py <argument>
```

The argument can be "startup" to install itself and run as a daemon. Also, it can be "delstartup" to remove itself and disable the daemon. The daemon installation will be fully automated, and will, by default, be configured to run during the deployment phase using the *Autoconfig tool*. The daemon will have the ability to identify the underlying operating system, be it Linux or Windows. The difference in behavior is described below:

- In the case of Linux OS:

  1. The path of the agents OSSEC directory will be set to */var/ossec*.
  2. The daemon will use the systemd to daemonize itself and run upon startup. The name of the daemon will be *soc_plugin_system.service*.
  3. The resulting logs of the execution will be sent to Wazuh using Syslog and will be in the following format:

     MON DAY HH:MM:SS HOSTNAME USERNAME: SOC_NES: PLUGIN_NAME: MSG

- In the case of Windows OS:

  1. The path of the agents OSSEC directory will be set to *C:\Program Files (x86)\ossec-agent*.
  2. The daemon will use the scheduled tasks (*schtasks*) to persist and run itself upon startup. The name of the scheduled task will be *SOC Plugin System*.
  3. The resulting logs of the execution will be logged at a file called "plugin_syslog.log", which is located within the OSSEC directory. The file will be monitored by the agent daemons by adding the configuration in Fig. 20 into *ossec.conf* and the log will be sent in the same Syslog format.

```
1   <localfile>
2       <location>plugin_syslog.log</location>
3       <log_format>syslog</log_format>
4   </localfile>
```

**Figure 20:** Windows Syslog File Configuration.

### 4.7.3. Retrieving New Plugins

To avoid wasting network resources, the daemon will not be polling the Wazuh manager and asking for its *[ID].json* and check on whether it has any updates; instead, the updates will be automatically pushed into the agents by utilizing the Wazuh daemons. The agent will continuously read its local shared directory contents (currently every 3 seconds), and whenever it spots a change, it will go through each plugin defined and compare the versions of the plugins to see if it must update any plugins or to pull new ones. Three cases are possible:

1. There is a totally new plugin: the plugin will run in a new process.

2. The new version is not equal to the old one: the old version process will be terminated, and a new version will run as a new process.

3. The plugin is deleted, where it is not intended to be used by the agent once more: the plugin's process will be terminated.

The *agentd* will be able to automatically get the Wazuh manager IP address from the *ossec.conf* file, and will be used to connect to the manager API. In order to run the new plugin, the daemon will fork a new process and pass the new plugin id and request the plugins from the API. The plugins will be shipped to agents in a zipped format; therefore, the agents will have to unzip the plugin and create the file structure locally.

### 4.7.4. Plugin Execution

. The daemon will use the Python's subprocess module to execute *script.py*. The subprocess will run the plugin code every "interval" defined in the *metadata.json* and capture its output. The script must print its output directly to the *console/stdout*. The output will be redirected by the agent daemon without any special communication between the two processes. Then, it will be analyzed and sent to the Wazuh manager. The output will exhibit the following form:

LOG: The Log That Should Be Sent To SOC
ARY: Arg1 Arg2 Arg3 ...

Where "LOG" is the logs that will be sent to Wazuh, using Syslog, which will be decoded to produce the corresponding alerts, and present them visually on the dashboards. The Active Response Yes (ARY) will be sent as a POST request to the API's /{plugin_id}/ar endpoint. The Arg1, Arg2 and Arg3 will be sent to the Wazuh manager in order to be passed to the active response code which will execute any needed countermeasures on the Wazuh manager. Here, the arguments' values must not contain any spaces. The Active Response No (ARN) is reserved for future use. At this point, the script will further execute any counter measures that are needed locally at the Wazuh agent machine to safeguard the machine against further threat.

### 4.7.5. PPIDPS Interaction Showcase

Consider Fig. 21, which shows the possible interactions in the API and how they affect the Wazuh manager and the agents.

Assume that the plugin system already possesses a plugin with an id of *0bab811d*, which is currently disabled.

The interactions will take place using the API's web interface, which is embedded into Kibana. The web interface will have a continually updated list of the installed plugins on the system via the */plugins/* API endpoint, where the SOC engineer can issue the desired changes and updates to those plugins or import new ones.

The effect of enabling the plugin with id *0bab811d* is shown in Blue. First, in **(1)**, using the web interface. The user will enable the interface, which will result in sending a request to the */plugins/0bab811d.json* to modify its *metadata.json*, changing its "enabled" property to "true".

In **(2)**, the local decoders and local alerts, which are part of the Analysis Engine, will be modified and appended by the ones defined in the plugin's *decoders.xml* and *rules.xml* files. Here, the active-response script will be pushed onto its appropriate place */active-response/plugins*, and be named by the id of the plugin; in this case, *0bab811d.py*. Lastly, the Wazuh manager service will be restarted, and the plugin is ready to use; hence, new *{ID}.json* files will be created at the shared directory – */shared/default/plugins* - for every agent id defined in the "agents" element of *metadata.json*. Finally, the Wazuh daemons will execute their tasks and distribute them to the agents instructing them to use the plugins.

The retrieval of the enabled plugin by an agent (id = 002) is shown in Green. The agent, responsible for monitoring any changes in its *002.json* file will spot the changes once the manager has pushed the new JSON files into the shared directory. This is shown in step **(1)**. In **(2)**, the agent will send a request – */plugins/0bab811d.zip?size=minimal*, to the API to download the plugin. In **(3)**, the API will respond by sending the plugin's script and metadata in a zipped format. In **(4)**, the agent will create a directory for the plugin, named pursuant with its id, and unzip the plugin inside the directory. In **(5)**, the delivered *script.py* will be executed by the *gentd* and consequently capture its output. The script will detect whether the system is being attacked by the type of attack that it is intended to detect and execute any possible remediation actions to salvage the system. Lastly, in **(6)**, the resulting output from executing the script will be sent by the *agentd* to the Wazuh manager. The LOG will be sent to the analysis engine for further analysis and, in the process, triggering alerts that will be posted to the dashboards. The ARY will be sent through a request – */0bab811d/ar* to the API, which will, in turn, pass it to the active response script and execute it (this is not shown in the figure to avoid making it messy) accordingly.

A SOC engineer requesting to download the plugin with id *0bab811d*, can accomplish that using the web interface. This is done perhaps to modify the script or add new decoders and rules. This is illustrated in purple. The web interface will send a request – */plugins/0bab811d.zip?size=full* to the API, which will cause the system to send the entire plugin contents, in zipped format, as shown in **(1)**. It then sends it to the SOC engineer for download, as shown in **(2)**.

In the figure, a SOC engineer, as marked in RED, has implemented a new plugin which he wants to import to the system with id of *3cd6bbc9*, using the web interface. The plugin would be imported as a zip file, which will cause a request – */plugins/* to the API with the zipped file as data. The Wazuh manager will receive the zipped plugin, create a directory using its id – *3cd6bbc9*, unzip it to the same directory location, and create the proper file structure, as shown in **(1)** and **(2)**. If the plugin is enabled, the same process described earlier (Blue) will take place.

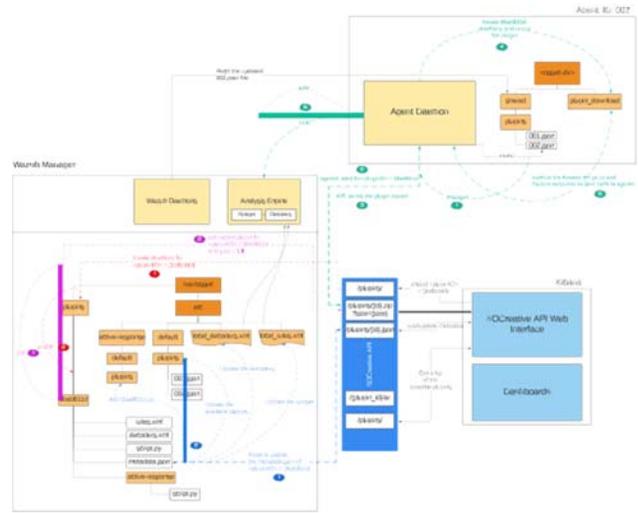

**Figure 21:** The PPIDPS Workflow.

### 4.7.6. Essence of The Design

As is readily evident, the design, and, hence, the implementation, is very agile, cohesive, and is platform independent. Further, the system can be expanded to work with more scripting languages easily; something commonly done to meet the needs of different security engineers. This allows for exchange of the plugins between the various SOC teams to be a very simple and seamless process, where a SOC team can receive a set of plugins and are able to modify them, including those of the triggered alerts' levels; this will adapt to varying risk profiles and needs, as the plugin continues to execute! In fact, this is particularly part of our plans to extend the plugin system with a web store, which we call a Security Operations Center Intrusion Detection and Prevention Plugins Store (SOC-IDPPS), which would contain hundreds of plugins that can detect and prevent possible threats where teams can also share plugins for various use cases; and, hence, enhance the security for any kind of organization to meet its risk profiles. The store will be fully embedded into the Wazuh web interface as well, making it an easy task to grab a plugin with a single mouse click.

## 5. Testing and Evaluation

This section will go through the process of running the *Autoconfig tool*, providing the needed input and testing the deployed environment against simple attacks to show that the environment is operating effectively. Also, we will assemble (design) a simple plugin on the PPIDPS in order to detect sophisticated attacks that may rely on zero-day vulnerabilities.

The environment we use in the testing is deployed in the Amazon Web Services (AWS) cloud and is shown in Fig. 22.

The same environment is used in the tests conducted in all subsequent subsections. The environment is a distributed Wazuh deployment, where we have a separate server for each component; a Wazuh Manager, an Elasticsearch, and the Kibana component. The environment also includes a limited set of two agents to test the effectiveness of our tools against both Linux and Windows 10 machines. The last component is a CISCO[1] router that runs locally as a virtual machine, which will also be used to test the auto-configuration process.

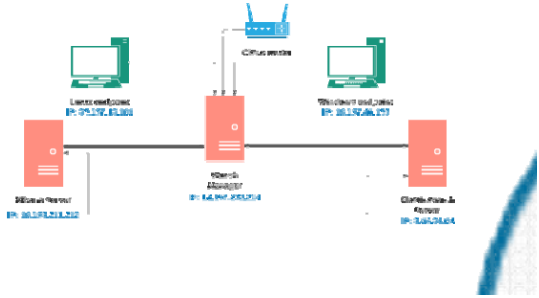

**Figure 22:** SOC Environment Topology in The Cloud.

In each test, we will mimic simple attack scenarios against the Linux and Windows agents to demonstrate associated alert firing responses on Wazuh.

### 5.1. Deploying the Clusters and Agents

First, and as precursor to deploying the system, the following must be manually done/known:

- The SSH daemon must be installed and enabled on all the machines (servers and agents) and configured to accept SSH keys login.

- All the SSH keys must be gathered and added to the */creds* directory of the Automation Engine file-structure.

- Python3 must be installed on all the machines. It is installed by default on the latest Linux machines; however, it must be installed manually on Windows agent machines and outdated Linux machines.

- The user should be aware of the IP addresses of the machines that will be deployed.

- The user who will use the *Autoconfig tool* needs to have Ansible packages installed prior to using the tool.

At this point, we would be ready to run the formatting tool – *formatter.py* to create the applicable formatting to be fed into the *Autoconfig tool*.

As shown in Fig. 23a, the required IPs along with the SSH users and SSH key paths (hidden) were inserted into the tool that will result in the creation of the *topology.txt*, which would readily include the information in encrypted and proper formats. Following this, we would be ready to run the *Autoconfig tool*, which will read the *topology.txt* file to initiate the deployment process. The deployment will start with Elasticsearch server installation and configuration as shown in Fig. 23b. This will be followed by the Kibana server installation and configuration as shown in Fig. 23c, and finally the deployment of the Wazuh Manager server as shown in Fig. 23d.

With the deployment of Wazuh and configuration of its components, the tool will ask the users whether they would be interested in integrating the system with Slack, as a ticketing system, and with VirusTotal, respectively. Fig. 24 illustrates the process of approval to add both integrations.

### 5.2. Autoconfig Tool Performance

Every component in the SOC will be handled by the appropriate Ansible-Playbook. Since the implemented playbooks are commonly designed with a level of intelligence to differentiate between the different platforms they run on, e.g., prior agent installation, the tool can readily determine whether the OS is Windows or Linux. Furthermore, the tool will also differentiate between various Linux distributions, e.g., CentOS, RedHat, Ubuntu, etc. Moreover, the tool will have a total error probability of zero opposite the deployment and configuration operations. Nonetheless, the tool is usually prone to human typing errors, since the *formatter.py* script often requires manual input from users. The entire SOC deployment and configuration processes during the testing stage, have been found to consume approximately 12 minutes towards a successful completion. The 12-minute period mentioned herein includes the time to download the packages/tools from their source, time to install them, and finally time needed to configure them. Hence, the time performance of the tool is directly affected by the network bandwidth and the endpoints/servers' specs.

We, also, illustrate the ease with which system scaling is accomplished; for instance, through deployment of two more Linux agents, we ran the formatter tool with the **-a** option for the purpose of adding agents and running the *Autoconfig tool* to deploy them. Fig. 25 reveals that the total number of active agents manifested amounts to four as reflected on the Wazuh web interface.

### 5.3. Testing Against SSH Login - Linux OS

In this subsection, we demonstrate system response to an attempted login leveraging SSH, as shown in Fig. 26, to a nonexistent 'test' user on a Linux- based platform. As shown in Fig. 27, an alert with level 5 is generated on the Wazuh manager, which reveals the time the attempt was initiated, by which user, together with its place of origin.

### 5.4. Testing Against Backdoor - Windows OS

In this subsection, we test against a Backdoor invasion on a Windows platform by executing a malicious file. As soon as a Trojan or a Backdoor attack is executed, it will add itself with the name *winhelp32.exe* to the following path:

C:\ProgramData\Microsoft\Windows\Start    Menu\Programs\Startup

Any executable in this path will be executed following a user's log in. The trojan would immediately need to root itself in the system. The special thing about this path is that it does not require a privilege to write any executable file

---

[1]We found that the predefined set of decoders delivered with Wazuh does not fully cover the different Cisco IOS log formats. Therefore, we had to manually alter these decoders to make the logs decoded and tested against the rule-set.

(a) Formatter Tool.

(b) Elastic Search Deployment.

(c) Kibana Deployment.

(d) Wazuh Server Deployment.

**Figure 23:** Deploying Wazuh Automatically for Formatter Tool, Elastic Search Deployment, Kibana Deployment, and Wazuh Server Deployment

**Figure 24:** Adding Integrations to the SOC Stack.

**Figure 25:** Agents Successfully Added - Main Interface.

**Figure 26:** SSH Attempt.

**Figure 27:** Alert for SSH Login Fail on Linux Agent.

ing a level 5 alert.

**Figure 28:** Alert for Integrity Violation on Windows Agent.

### 5.5. VirusTotal Integration Test

The previously tested executable is a slightly mutated version of the Bifrose Backdoor trojan citeBackdoor42:online. The Windows Defender Antivirus, alone, failed to detect this version of the malware when it was downloaded. But since we had added the VirusTotal integration to the SOC stack, any FIM triggered alert will also cause the modified files to be passed to and scanned by VirusTotal. The mutated malware was detected by 45 Antivirus engines on VirusTotal.

on it. For that reason, the trojan will use this directory to be launched unto the system. For this reason, this folder was inherently set to be monitored in real-time by the File Integrity Monitoring (FIM) module in the Wazuh Windows agent during the configuration and deployment phase. Fig. 28 shows that Wazuh instantly detects the file-system changes trigger-This is illustrated in Fig. 29. From the figure, a detailed alert manifestation is evident opposite the malware caught.

**Figure 29:** Alert due to Malware Detection by VirusTotal.

## 5.6. Testing Against Configuration Changes - Cisco Router

To mimic a test scenario against the Cisco router, we had created an SSH login with the name *admin*. Any configuration attempt by the new user (and any other user), using the command *"configure terminal"* to access the Global Configuration Mode on Cisco IOS, will result in sending a log that triggers an alert on Wazuh. Following a successful SSH login to the Cisco router an alert is generated and is shown in Fig. 30

**Figure 30:** Alert Generated due to Configurations Change Attempt.

## 5.7. Real-World Attacks Test

It has been found that our IP addresses for our Amazon AWS servers were being SSH brute-forced by distributed bots. Fig. 31 shows the total number of alerts generated due to brute-force attacks.

**Figure 31:** Total Authentication Failure Alerts.

## 5.8. Ticketing System Test

Upon adding the Slack integration, the *Autoconfig tool* is used to configure the system to issue an alert for any level-5 threat, or above, automatically to Slack. These alerts will be further investigated by SOC analysts, and where they are able to identify any suspicious irregularities, they would readily be able to instantly open a ticket directly with the resident engineers at the organization. As a result of our test, a ticket created for an alert previously is pushed to Slack, where the SOC analyst reported a need for further investigation. The ticket, as assigned to a SOC analyst, was issued with 'Open' status, which would readily be changed to 'Closed' upon resolving the issue.

## 5.9. Programmable Plugin-based IDPS Test

Our plugin architecture serves as an extra layer of protection between the kernel layer and the user-space layer. It will hook and manipulate the system to identify the "read" operations on specific files. In our case, the plugin is programmed to generate alerts each time the PASSWD file is accessed successfully. Furthermore, it is programmed to create an alert and trigger the active response if the SHADOW file is accessed. In connection with the active response mechanism, it is programmed to send emails reporting the security event that took place. The plugin can easily be modified to be activated on the "writing" operation or even on querying the file's size or name.

To test our programmable plugin-based IDPS, we started out by importing the *ZeroDayFileWatch* plugin into our SOC. After the plugin is imported and its status is enabled, we, also, considered mimicking the "*vsftpd*" vulnerability as a zero-day vulnerability. We used Kali Linux and Metasploit to perform our attack by setting the required configuration using Kali and Metasploit. The vulnerability being exploited is illustrated in Fig. 32a. Following the read operation of the PASSWD file, leveraging the cat command, an alert is immediately generated as shown in Fig. 32b. This alert serves as an indication of system intrusion. However, the plugin is programmed just to set an alert of level 10 if the PASSWD file is read, and in the case of a shadow file read, it will set an alert with level 15 and activate the active-response code. The active-response code will, in turn, generate an alert and send an email to the SOC people as shown in Figs. 32c and 32d.

It is worthy of noting, however, that in our test case, the plugin was designed to detect intrusions based on any file system read operation executed on numerous sensitive files. Meanwhile, another design can be contrived: for instance, a plugin can detect intrusions commensurate with TCP/UDP connection tables for susceptible and monitored servers. This can be done by hooking the connection system calls; i.e., hooking the "connect", "accept", and "socket" kernel functions. If any of these functions are called under certain conditions, an alert or/and active response would readily take place.

## 6. Conclusion

Results presented in this paper have addressed the development of a low-cost, fully adaptive, plug-and-play, fully operating Security Operations Center (SOC). Automation is simply one part of a much-needed effort to shore up SOC effectiveness, and it certainly offers an essential component that can help make a big difference in the process. Consequently, automation for low-risk, high-yield tasks has been provided. Meanwhile, other parts have also been addressed with the purpose of contriving a new SOC design with the necessary software and tools onboard to achieve an adaptive security operations center. Building an adaptable system to meet any organizational deployment needs is considered the project's core goal. We have discussed in detail our proposed framework which is used to deploy the SOC and extend its components with powerful horizontal scalability techniques; all within few minutes and with no possibilities for human error. Additionally, we have demonstrated the adaptability in terms of budgeting aspects, so that any organization would be able to access it as an open-source security solution.

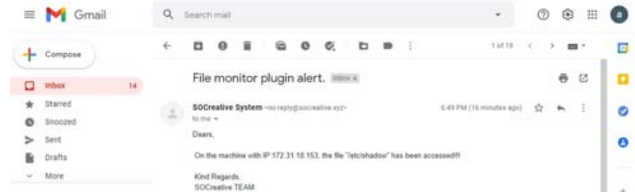

(a) Matasploit Exploit.

(b) PASSWD File Alert.

(c) Shadow File Alert.

(d) Active Response Email Notification.

**Figure 32:** Programmable Plugin-based IDPS Test Scenarios

Furthermore, to handle the diversity of organizational threats and risk-profiles, a one-of-a-kind feature, which we have dubbed a "Programmable Plugin-based Intrusion Detection and Prevention System (PPIDPS)" was introduced. Using the PPIDPS will further offer the ability to add any tool to the monitored devices while also getting logs that can trigger alerts when something goes wrong. In other words, the SOC is no longer a static module with fixed protection criteria. This gives an organization the ability to detect zero-day vulnerabilities and customize the SIEM. As such, the system is made fully programmable to meet the organizations and risk-profiles set by the security engineers. Moreover, the PPIDPS is offered in a plugin-based environment making it the first system of its kind in enabling risk-profile, and threat mitigations between SOCs across the globe all without limits. The system can activate a programmable active-response; whenever an anomaly or threat is detected, the active-response code will run on the Wazuh Manager device and execute any needed counter tasks. The plugin introduced is fully programmable; a security engineer can readily design the protection criteria that would execute the proper incident handling mechanism, as appropriate, where all that is done aboard the agent device itself.

As future work, We are planning establish an online platform called Security Operations Center - Intrusion Detection And Prevention Plugins Store (*SOC-IDPPS*), where the SOC security engineers and analysts can push their custom plugins, which they frequently use with the implemented (*PPIDPS*). These plugins can be added to Wazuh on the fly through the web interface, where the users can see the updated and newly added plugins as they are pushed. Furthermore, these plugins will be tested to ensure their functionality and rated and reviewed by the community.